\documentclass[12pt,aps,onecolumn,superscriptaddress]{revtex4}
\usepackage{graphicx}
\usepackage{epsfig}
\usepackage{amssymb,amsmath}

\begin{document}

\title{
Measurement of the $e^+e^-\to\pi^0\gamma$ cross section in the energy range
from 1.08 to 2.0~GeV with the SND detector
}

\author{M.~N.~Achasov}
\affiliation{Budker Institute of Nuclear Physics, SB RAS, Novosibirsk, 630090, Russia}
\affiliation{Novosibirsk State University, Novosibirsk, 630090, Russia}
\author{A.~E.~Alizzi}
\affiliation{Budker Institute of Nuclear Physics, SB RAS, Novosibirsk, 630090, Russia}
\affiliation{Novosibirsk State University, Novosibirsk, 630090, Russia}
\author{A.~Yu.~Barnyakov}
\affiliation{Budker Institute of Nuclear Physics, SB RAS, Novosibirsk, 630090, Russia}
\author{K.~I.~Beloborodov}
\affiliation{Budker Institute of Nuclear Physics, SB RAS, Novosibirsk, 630090, Russia}
\affiliation{Novosibirsk State University, Novosibirsk, 630090, Russia}
\author{A.~V.~Berdyugin}
\affiliation{Budker Institute of Nuclear Physics, SB RAS, Novosibirsk, 630090, Russia}
\affiliation{Novosibirsk State University, Novosibirsk, 630090, Russia}
\author{D.~E.~Berkaev}
\affiliation{Budker Institute of Nuclear Physics, SB RAS, Novosibirsk, 630090, Russia}
\affiliation{Novosibirsk State University, Novosibirsk, 630090, Russia}
\author{A.~G.~Bogdanchikov}
\affiliation{Budker Institute of Nuclear Physics, SB RAS, Novosibirsk, 630090, Russia}
\author{A.~A.~Botov}
\affiliation{Budker Institute of Nuclear Physics, SB RAS, Novosibirsk, 630090, Russia}
\author{V.~S.~Denisov}
\affiliation{Budker Institute of Nuclear Physics, SB RAS, Novosibirsk, 630090, Russia}
\author{T.~V.~Dimova}
\affiliation{Budker Institute of Nuclear Physics, SB RAS, Novosibirsk, 630090, Russia}
\affiliation{Novosibirsk State University, Novosibirsk, 630090, Russia}
\author{V.~P.~Druzhinin}
\affiliation{Budker Institute of Nuclear Physics, SB RAS, Novosibirsk, 630090, Russia}
\affiliation{Novosibirsk State University, Novosibirsk, 630090, Russia}
\author{R.~A.~Efremov}
\affiliation{Budker Institute of Nuclear Physics, SB RAS, Novosibirsk, 630090, Russia}
\affiliation{Novosibirsk State Technical University, Novosibirsk, 630073, Russia}
\author{E.~A.~Eminov}
\affiliation{Budker Institute of Nuclear Physics, SB RAS, Novosibirsk, 630090, Russia}
\author{L.~V.~Kardapoltsev}
\affiliation{Budker Institute of Nuclear Physics, SB RAS, Novosibirsk, 630090, Russia}
\affiliation{Novosibirsk State University, Novosibirsk, 630090, Russia}
\author{A.~A.~Kattsin}
\affiliation{Budker Institute of Nuclear Physics, SB RAS, Novosibirsk, 630090, Russia}
\author{V.~R.~Khamidullin}
\affiliation{Budker Institute of Nuclear Physics, SB RAS, Novosibirsk, 630090, Russia}
\affiliation{Novosibirsk State University, Novosibirsk, 630090, Russia}
\author{A.~G.~Kharlamov}
\affiliation{Budker Institute of Nuclear Physics, SB RAS, Novosibirsk, 630090, Russia}
\affiliation{Novosibirsk State University, Novosibirsk, 630090, Russia}
\author{I.~A.~Koop}
\affiliation{Budker Institute of Nuclear Physics, SB RAS, Novosibirsk, 630090, Russia}
\affiliation{Novosibirsk State University, Novosibirsk, 630090, Russia}
\author{A.~A.~Korol}
\affiliation{Budker Institute of Nuclear Physics, SB RAS, Novosibirsk, 630090, Russia}
\affiliation{Novosibirsk State University, Novosibirsk, 630090, Russia}
\author{D.~P.~Kovrizhin}
\affiliation{Budker Institute of Nuclear Physics, SB RAS, Novosibirsk, 630090, Russia}
\author{A.~P.~Kryukov}
\affiliation{Budker Institute of Nuclear Physics, SB RAS, Novosibirsk, 630090, Russia}
\author{A.~S.~Kupich}
\affiliation{Budker Institute of Nuclear Physics, SB RAS, Novosibirsk, 630090, Russia}
\affiliation{Novosibirsk State University, Novosibirsk, 630090, Russia}
\author{A.~R.~Mamedov}
\affiliation{Budker Institute of Nuclear Physics, SB RAS, Novosibirsk, 630090, Russia}
\affiliation{Novosibirsk State University, Novosibirsk, 630090, Russia}
\author{N.~A.~Melnikova}
\affiliation{Budker Institute of Nuclear Physics, SB RAS, Novosibirsk, 630090, Russia}
\author{N.~Yu.~Muchnoi}
\affiliation{Budker Institute of Nuclear Physics, SB RAS, Novosibirsk, 630090, Russia}
\affiliation{Novosibirsk State University, Novosibirsk, 630090, Russia}
\author{A.~E.~Obrazovsky}
\affiliation{Budker Institute of Nuclear Physics, SB RAS, Novosibirsk, 630090, Russia}
\author{A.~A.~Oorzhak}
\email{A.A.Oorzhak@inp.nsk.su}
\affiliation{Budker Institute of Nuclear Physics, SB RAS, Novosibirsk, 630090, Russia}
\affiliation{Novosibirsk State University, Novosibirsk, 630090, Russia}
\author{I.~V.~Ovtin}
\affiliation{Budker Institute of Nuclear Physics, SB RAS, Novosibirsk, 630090, Russia}
\affiliation{Novosibirsk State University, Novosibirsk, 630090, Russia}
\author{E.~V.~Pakhtusova}
\affiliation{Budker Institute of Nuclear Physics, SB RAS, Novosibirsk, 630090, Russia}
\author{E.~A.~Perevedentsev}
\affiliation{Budker Institute of Nuclear Physics, SB RAS, Novosibirsk, 630090, Russia}
\affiliation{Novosibirsk State University, Novosibirsk, 630090, Russia}
\author{I.~A.~Polomoshnov}
\affiliation{Budker Institute of Nuclear Physics, SB RAS, Novosibirsk, 630090, Russia}
\affiliation{Novosibirsk State University, Novosibirsk, 630090, Russia}
\author{K.~V.~Pugachev}
\affiliation{Budker Institute of Nuclear Physics, SB RAS, Novosibirsk, 630090, Russia}
\affiliation{Novosibirsk State University, Novosibirsk, 630090, Russia}
\author{Yu.~A.~Rogovsky}
\affiliation{Budker Institute of Nuclear Physics, SB RAS, Novosibirsk, 630090, Russia}
\affiliation{Novosibirsk State University, Novosibirsk, 630090, Russia}
\author{V.~A.~Romanov}
\affiliation{Budker Institute of Nuclear Physics, SB RAS, Novosibirsk, 630090, Russia}
\affiliation{Novosibirsk State University, Novosibirsk, 630090, Russia}
\author{A.~I.~Senchenko}
\affiliation{Budker Institute of Nuclear Physics, SB RAS, Novosibirsk, 630090, Russia}
\author{S.~I.~Serednyakov}
\affiliation{Budker Institute of Nuclear Physics, SB RAS, Novosibirsk, 630090, Russia}
\affiliation{Novosibirsk State University, Novosibirsk, 630090, Russia}
\author{Yu.~M.~Shatunov}
\affiliation{Budker Institute of Nuclear Physics, SB RAS, Novosibirsk, 630090, Russia}
\author{D.~A.~Shtol}
\affiliation{Budker Institute of Nuclear Physics, SB RAS, Novosibirsk, 630090, Russia}
\author{Z.~K.~Silagadze}
\affiliation{Budker Institute of Nuclear Physics, SB RAS, Novosibirsk, 630090, Russia}
\affiliation{Novosibirsk State University, Novosibirsk, 630090, Russia}
\author{K.~D.~Sungurov}
\affiliation{Budker Institute of Nuclear Physics, SB RAS, Novosibirsk, 630090, Russia}
\affiliation{Novosibirsk State University, Novosibirsk, 630090, Russia}
\author{I.~K.~Surin}
\affiliation{Budker Institute of Nuclear Physics, SB RAS, Novosibirsk, 630090, Russia}
\author{M.~V.~Timoshenko}
\affiliation{Budker Institute of Nuclear Physics, SB RAS, Novosibirsk, 630090, Russia}
\author{Yu.~V.~Usov}
\affiliation{Budker Institute of Nuclear Physics, SB RAS, Novosibirsk, 630090, Russia}
\author{I.~M.~Zemlyansky}
\affiliation{Budker Institute of Nuclear Physics, SB RAS, Novosibirsk, 630090, Russia}
\author{V.~N.~Zhabin}
\affiliation{Budker Institute of Nuclear Physics, SB RAS, Novosibirsk, 630090, Russia}
\affiliation{Novosibirsk State University, Novosibirsk, 630090, Russia}
\author{V.~V.~Zhulanov}
\affiliation{Budker Institute of Nuclear Physics, SB RAS, Novosibirsk, 630090, Russia}
\affiliation{Novosibirsk State University, Novosibirsk, 630090, Russia}

\begin{abstract}
In the experiment with the SND detector at the VEPP-2000 $e^+e^-$
collider, the cross section for the process $e^+e^-\to\pi^0\gamma$ has
been measured in the center-of-mass energy range
$\sqrt{s}=1.08$--2.0~GeV. The analysis is based on data with an
integrated luminosity of 745~pb$^{-1}$, accumulated in 2017--2024.
The obtained cross section is consistent with previous measurements
but has significantly better accuracy. The cross section is
well described by the vector meson dominance model, which takes into
account the contribution of excited resonances of the $\rho$ and
$\omega$ families. The hypothesis of no excited-resonance contribution
is excluded at the $4.4\sigma$ level.
\end{abstract}

\maketitle

\section{Introduction}

Radiative transitions between light vector mesons of the $\rho$,
$\omega$, and $\phi$ families and the pseudoscalar mesons $\pi^0$ and
$\eta$ have traditionally been used to test various phenomenological
models of low-energy hadron
interactions~\cite{landsberg1985electromagnetic, o1981radiative, o1997rho}.

In the energy range from 1.05 to 2.0~GeV, experimental data indicate a
high density of excited states of light vector mesons:
$\omega(1420)$, $\rho(1450)$, $\omega(1650)$, $\phi(1680)$, and
$\rho(1700)$~\cite{pdg2026}. Measurement of the radiative decay widths
of these resonances makes it possible to study their properties.
In addition to ordinary $q\bar{q}$ mesons, exotic states such as hybrid mesons
($q\bar{q}g$) are expected in this energy
range~\cite{isgur1985gluonic, close1995page}. Since hybrid states can
mix with ordinary $q\bar{q}$ states, their identification is a
challenging experimental task that requires the analysis of all possible
decay modes. Radiative decays, whose branching fractions are relatively
well predicted in the quark model, may play an important role in the
search for hybrid vector states~\cite{close2002radiative}. Besides
hybrid mesons, manifestations of multiquark states~\cite{jaffe1977} and
other exotic particles~\cite{pdg2026} are expected.

Experimental data on radiative processes are also used to develop
phenomenological models of meson-photon transition form factors.
These form factors play an important role in the calculation of the
hadronic light-by-light scattering contribution to the anomalous
magnetic moment of the muon~\cite{aoyama2020, aliberti2025, Hoferichter2018}.

The cross section for the process $e^+e^-\to\pi^0\gamma$ was measured in
the SND experiments~\cite{achasov2016study, achasov2018measurement, achasov2025measurement}
at the VEPP-2M and VEPP-2000 colliders, and in the CMD-2 
experiment~\cite{akhmetshin2005study} at the VEPP-2M collider. The most accurate
measurement of the $e^+e^-\to\pi^0\gamma$ cross section in the center-of-mass (c.m.) energy range $\sqrt{s}=1.075$--2.0~GeV was performed in the SND experiment at 
VEPP-2000~\cite{achasov2018measurement}. That measurement was based on data with an 
integrated luminosity of 41~pb$^{-1}$, collected in 2010--2012.

In this paper, we present a measurement of the $e^+e^-\to\pi^0\gamma$
cross section in the energy range from 1.08 to 2.0~GeV based on data with
an integrated luminosity of about 745~pb$^{-1}$, collected from 2017
to 2024 by the SND detector at the VEPP-2000 $e^+e^-$
collider~\cite{shatunov2016status}. 
\section{Detector and experiment}

SND (Spherical Neutral
Detector)~\cite{achasov2000spherical, achasov2009first, aulchenko2009snd,
barnyakov2009high, aulchenko2009daq}
is a general-purpose non-magnetic detector operating at the VEPP-2000
$e^+e^-$ collider in the energy range
$\sqrt{s}=0.32$--2.0~GeV~\cite{timoshenko2020status}. The SND detector
consists of the following subsystems: a tracking system, a system of
aerogel Cherenkov counters, an electromagnetic calorimeter, and a muon
system. The tracking system consists of a nine-layer drift chamber and
a single-layer proportional chamber in a common gas volume. The aerogel
Cherenkov counters with a refractive index of $n=1.13$, located around
the tracking system, are used to identify charged kaons.

The most important SND subsystem for the analysis described in this
paper is the electromagnetic calorimeter. The calorimeter is spherical
and consists of three layers. It contains 1640 counters based on NaI(Tl)
crystals. The calorimeter covers polar angles from $18^\circ$
to $162^\circ$, which corresponds to a solid angle of 95\% of $4\pi$.
In this analysis, the calorimeter is used to measure the energies and
angles of photons. A reconstructed photon is defined as a cluster of hit
calorimeter crystals with an energy deposition exceeding 20~MeV that is
not associated with any charged track. The spherical shape of the
calorimeter provides uniform sensitivity to photons over the solid
angle. The energy resolution of the calorimeter is
$\sigma_E/E=4.2\%/\sqrt[4]{E({\rm GeV})}$, and the angular resolution
is about $1.5^\circ$.

The Monte Carlo event generators for the signal and background hadronic
processes include radiative corrections~\cite{kuraev1985radiative}.
The angular distribution of the hard photon emitted from the initial
state is generated according to Ref.~\cite{bonneau1971hard}.
The interaction of the generated particles with the detector material
is simulated using the GEANT4 package~\cite{agostinelli2003, allison2016}. The simulation
takes into account variations in the detector conditions during data
taking, for example, dead counters in the calorimeter. 
To account for the overlap of the beam-generated background with the
events under study, the simulation uses special background events
recorded during data taking with a random trigger. These events are
overlaid on the simulated events, thereby reproducing the experimental
events with the additional background tracks and photons.

This analysis is based on data with an integrated luminosity of
745~pb$^{-1}$, recorded with the SND detector from 2017 to 2024 in the
energy range $\sqrt{s}=1.08$--2.0~GeV. The data were collected at 153
energy points. Since the $e^+e^-\to\pi^0\gamma$ cross section
is small, the energy points are combined into intervals and
the numbers of selected events and the integrated luminosities are
summed within each interval.

The integrated luminosity ($\mathrm{IL}$) is measured
using the process $e^+e^-\to\gamma\gamma$. The event selection and 
the estimate of the systematic uncertainty in the luminosity 
follow Ref.~\cite{achasov2024measurement}.
Compared with that work, we add the conditions
$E_{\mathrm{tot}}>0.65\sqrt{s}$ and $P_{\mathrm{tot}}<0.3\sqrt{s}$,
which are used for selection of $\pi^0\gamma$ events in Sec.~\ref{sel}.
Here $E_{\mathrm{tot}}$ is the total energy deposition in the
calorimeter, and $P_{\mathrm{tot}}$ is the total momentum of the event
calculated from the energy depositions in the calorimeter crystals.
We also require that $N_\gamma=2$, where $N_\gamma$ is the number of
reconstructed photons in the event. This requirement corresponds
to $N_\gamma=3$ in the selection of $\pi^0\gamma$ events.
Using the process
$e^+e^-\to\gamma\gamma$ for normalization makes it possible to cancel
some systematic uncertainties in the measurement of the 
$e^+e^-\to\pi^0\gamma$ cross section. These include the uncertainties
associated with the event selection in the hardware first-level
trigger, the requirements of no charged tracks and no extra photons in 
the event, and the conditions on the parameters $E_{\mathrm{tot}}$ and
$P_{\mathrm{tot}}$. The systematic uncertainty associated with
the other selection criteria is estimated to be about 1\%.
The measured integrated luminosity is listed in Table~\ref{table}.
\section{Event selection\label{sel}}

The selection of $e^+e^-\to\pi^0\gamma$ events is performed in two
stages. At the first stage, we select events with exactly three 
photons with energies above 50~MeV and no charged tracks, and
satisfying the conditions $E_{\mathrm{tot}}>0.65\sqrt{s}$ and 
$P_{\mathrm{tot}}<0.3\sqrt{s}$.

For events that pass these conditions, a kinematic fit to the
hypothesis $e^+e^-\to3\gamma$ is performed with the requirement of
energy and momentum conservation. For further analysis, we select
events with $\chi^2_{3\gamma}<30$, in which one of the two-photon 
invariant masses lies in the range $70<M_{\gamma\gamma}<200$~MeV,
where $\chi^2_{3\gamma}$ is the kinematic fit $\chi^2$.
In the energy range under study, the
$M_{\gamma\gamma}$ value chosen in this way corresponds to the mass 
recoiling against the most energetic photon. 
At this stage, the two-photon invariant
masses and the other selection parameters are calculated using the
photon angles and energies refined by the kinematic fit.
The use of tight selection conditions $N_\gamma=3$ and
$\chi^2_{3\gamma}<30$ allows the multiphoton hadronic background to 
be suppressed by a factor of about 20, while the signal detection
efficiency decreases by 12\%.

The dominant background for this analysis comes from $e^+e^-$
annihilation into three photons.
The backgrounds from annihilation into four or more photons and into
two photons with an additional spurious photon must also be taken
into account. In the latter case, the additional photon arises from
splitting of the electromagnetic shower or from beam-generated
background.
These processes are simulated with the BabaYaga@NLO Monte
Carlo generator~\cite{balossini2006matching, balossini2008photon}. It uses exact matrix
elements for $e^+e^-\to2\gamma$ and $3\gamma$. In addition, the emission
of an arbitrary number of extra photons is generated using the QED
parton shower algorithm.

A small contribution to the background also comes from the hadronic
processes $e^+e^-\to\pi^0\pi^0\gamma$ and $e^+e^-\to\eta\gamma\gamma_{\rm ISR}$,
where $\gamma_{\rm ISR}$ is a photon emitted from the initial state.
The contribution of other hadronic processes is studied using a
generator~\cite{generator2026} that includes all known hadronic
processes in the energy region under study. This contribution is found
to be negligible.

\begin{figure}[!t]
\centering
\includegraphics[width=0.5\textwidth]{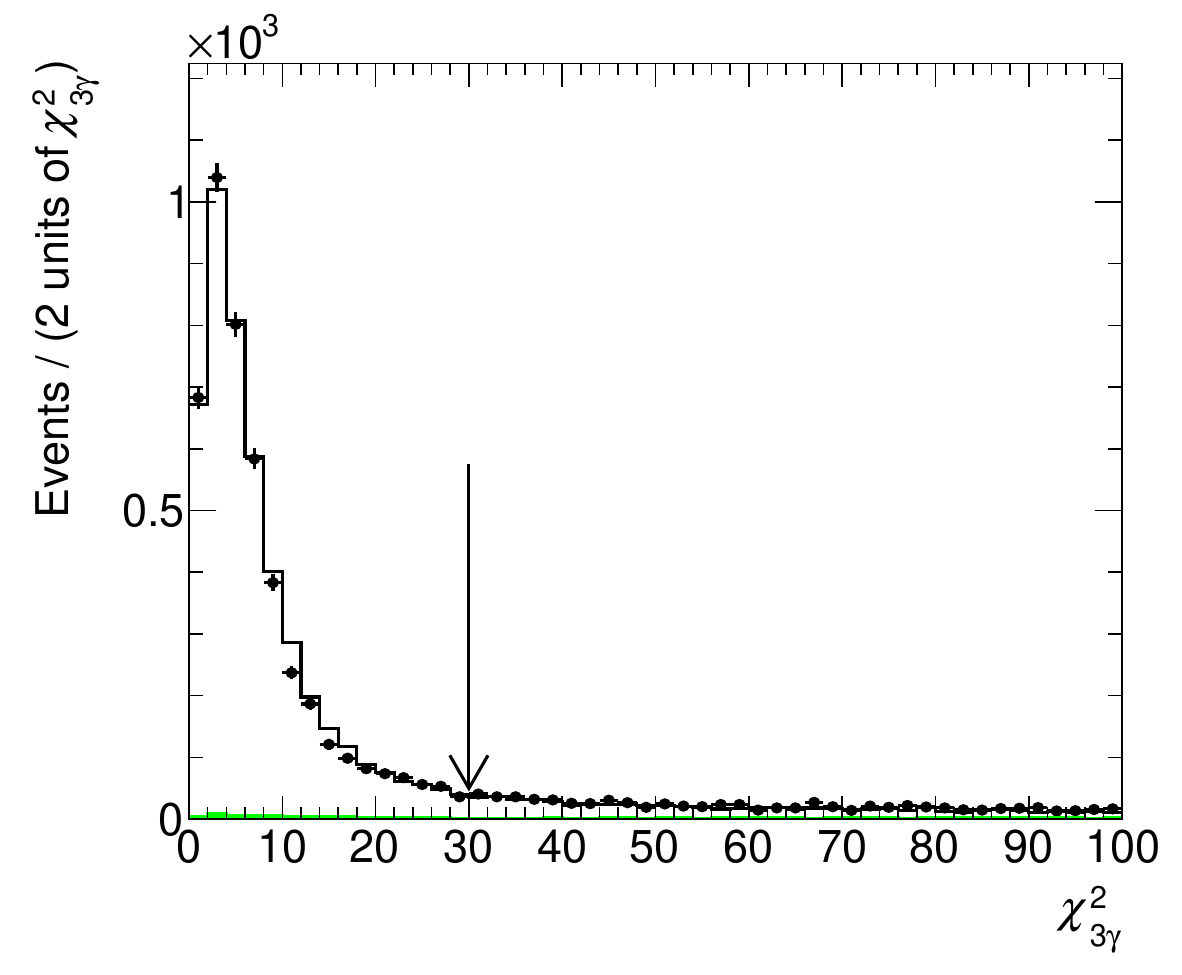}
\caption{The $\chi^2_{3\gamma}$ distribution at $\sqrt{s}=1.282$~GeV for
data (points with error bars) and simulation of the signal and
background processes (solid histogram). The shaded (green) histogram shows the
contribution of the multiphoton hadronic background. The simulated
distributions are normalized to the expected number of events. The
arrow indicates the condition $\chi^2_{3\gamma}<30$.}
\label{fig:chi2}
\end{figure}

The $\chi^2_{3\gamma}$ distribution for data and simulation at
$\sqrt{s}=1.282$~GeV is shown in Fig.~\ref{fig:chi2}. The distribution is dominated by
the background from the process $e^+e^-\to3\gamma$, whose distribution
is close to that of the signal process. In the region
$\chi^2_{3\gamma}<30$, the fraction of signal events is about 5\%.
The background in this region consists of $e^+e^-\to3\gamma$ (98.5\%),
multiphoton hadronic processes (1.2\%), and
$e^+e^-\to\pi^0\gamma\gamma_{\rm ISR}$ (0.3\%). The last process is
dominated by the radiative return to the $\omega$ and $\phi$
resonances followed by the $\omega,\phi\to\pi^0\gamma$ decays.

To suppress the background from the process $e^+e^-\to3\gamma$,
conditions are imposed on the photon polar angles
($36^\circ<\theta_\gamma<144^\circ$) and on the minimum photon energy in
the event. The polar angle conditions ensure that the electromagnetic
showers from the photons are fully contained in the calorimeter.
The optimal condition on the minimum photon energy is chosen by
maximizing the expected signal significance. For
$1.08\leq\sqrt{s}<1.625$~GeV, the threshold
$E_{\gamma,\mathrm{min}}>125$~MeV is chosen, and for
$1.625\le\sqrt{s}\le2.0$~GeV, the threshold is
$E_{\gamma,\mathrm{min}}>150$~MeV.
\section{Fit of the $M_{\gamma\gamma}$ distribution\label{mrec}}

The number of selected $e^+e^-\to\pi^0\gamma$ events
($N_{\pi^0\gamma}$) is determined from a fit to the $M_{\gamma\gamma}$
spectrum by the sum of the signal ($F_{\mathrm{sig}}$) and background
($F_{\mathrm{bkg}}$) distributions. The signal distribution is described
by a sum of two Gaussian functions whose parameters are determined from
a fit to the simulated $M_{\gamma\gamma}$ spectrum. The simulation of
the process $e^+e^-\to\pi^0\gamma(\gamma)$ includes the emission of an
additional photon $\gamma_{\rm ISR}$ from the initial state. In particular, the radiative
return to the $\omega$ and $\phi$ resonances is simulated. To
correctly reproduce the fraction of such events, the event generator uses the
data on the $e^+e^-\to\pi^0\gamma$ Born cross section obtained in this
work and in Refs.~\cite{achasov2016study, achasov2025measurement}.
The $M_{\gamma\gamma}$ spectrum for simulated
$e^+e^-\to\pi^0\gamma(\gamma)$ events is divided into two parts
according to the energy of the photon $\gamma_{\rm ISR}$
($E_{\mathrm{ISR}}$). For $1.08\le\sqrt{s}\le1.125$~GeV,
the boundary photon energy is 50~MeV: events with
$E_{\mathrm{ISR}}>50$~MeV are treated as background. For
$\sqrt{s}>1.125$~GeV, the boundary energy is 100~MeV.

When fitting the data, the function $F_{\mathrm{sig}}$ obtained from
simulation is modified to account for a possible difference in the
signal shape between data and simulation. A common shift of the peak
position $\Delta M$ and an additional variance $\Delta\sigma^2$ are
introduced, which modify the parameters of the Gaussian functions
obtained from simulation:
$M^{\mathrm{data}}_{\gamma\gamma}=M^{\mathrm{MC}}_{\gamma\gamma}+\Delta M$
and $\sigma^2_{\mathrm{data}}=\sigma^2_{\mathrm{MC}}+\Delta\sigma^2$.
The values of $\Delta M$ and $\Delta\sigma^2$ are determined from a fit
to the data $M_{\gamma\gamma}$ distributions in the region
near the maximum of the $\phi(1020)$ resonance
($\sqrt{s}=1.019$--1.020~GeV). They are found to be $\Delta M=(-0.16\pm0.17)$~MeV
and $\Delta\sigma^2=(15.5\pm3.6)$~MeV$^2$. In the fit to the data, the
parameters $\Delta M$ and $\Delta\sigma^2$ are fixed at these values.

\begin{figure}[t]
\centering
\includegraphics[width=0.6\textwidth]{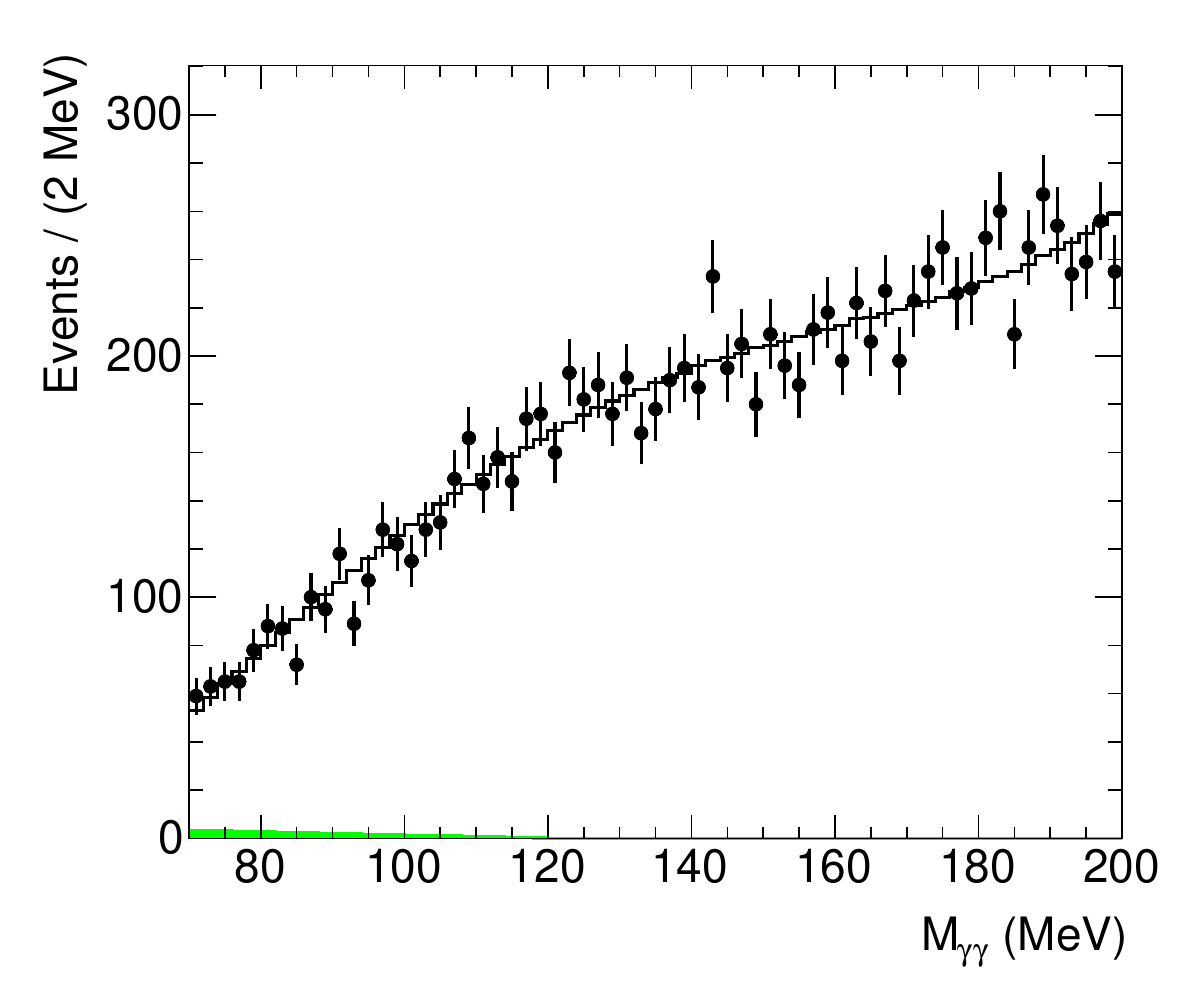}
\caption{The $M_{\gamma\gamma}$ distribution for data events (points with
error bars) at $\sqrt{s}=1.282$~GeV, selected with the condition that at
least one photon is detected outside the region
$36^\circ<\theta_\gamma<144^\circ$. The histogram is
the result of the fit described in the text.}
\label{fig:ggg}
\end{figure}

\begin{figure}[t]
\centering
\includegraphics[width=0.6\textwidth]{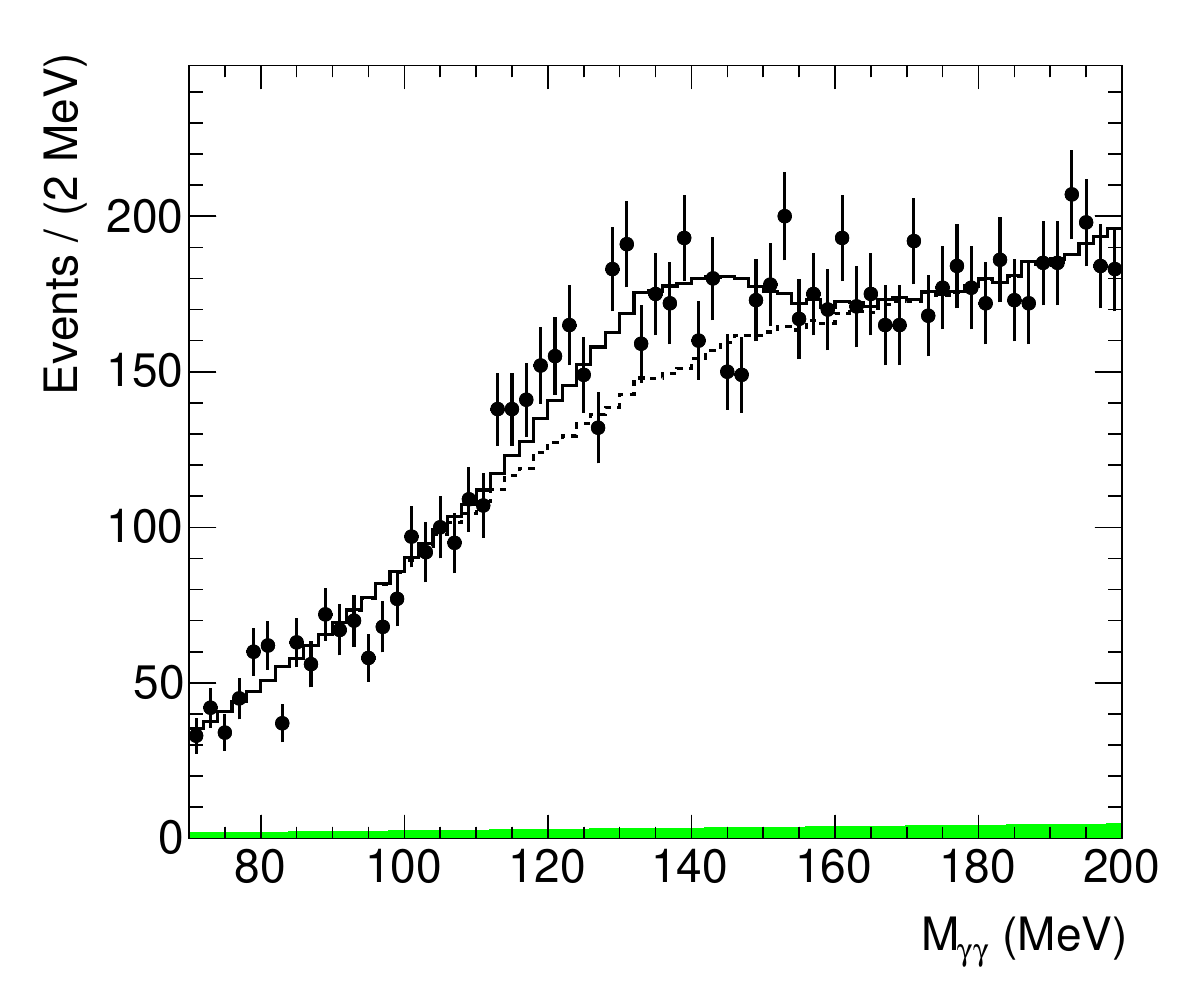}
\caption{The $M_{\gamma\gamma}$ distribution for data events (points with
error bars) at $\sqrt{s}=1.282$~GeV. The solid histogram is the result of
the fit described in the text. The dashed histogram represents the
background distribution. The green area shows the contribution of the
linear function.}
\label{fig:mgg}
\end{figure}

The shape of the $M_{\gamma\gamma}$ spectrum for the dominant background
process $e^+e^-\to3\gamma$ is described by a linear combination of
Chebyshev polynomials of the first kind up to fourth order, whose
coefficients are determined from a fit to the simulated spectrum.
Increasing the polynomial order above four does not improve the fit
quality. The contributions of the other background processes,
$e^+e^-\to\pi^0\pi^0\gamma$, $e^+e^-\to\pi^0\gamma\gamma_{\rm ISR}$, and
$e^+e^-\to\eta\gamma\gamma_{\rm ISR}$, are calculated using simulation. 
The total background function $F_{\mathrm{bkg}}$ is the sum of the
Chebyshev polynomials (for $e^+e^-\to3\gamma$), the simulated histograms
for the other background processes, and a linear function. 
The linear function compensates for a possible difference between data
and simulation in the shape and normalization of the background spectrum.
Since the statistics of the
$e^+e^-\to3\gamma$ simulation exceed the data statistics by only a
factor of 2--3, the statistical errors of the Chebyshev polynomial
coefficients determined from simulation lead to an uncertainty in the
background level comparable to the statistical errors of data.
To take them into
account properly, the simulated $M_{\gamma\gamma}$ distribution for the
process $e^+e^-\to3\gamma$ is fitted simultaneously with the
data distribution, with the Chebyshev polynomial coefficients
common to both spectra. Other free parameters of the simultaneous fit
are the number of signal events $N_{\pi^0\gamma}$ and the
parameters of the linear function.

Since the shape of the $M_{\gamma\gamma}$ spectrum for
$e^+e^-\to3\gamma$ events is not linear, it is necessary to verify that
the simulation reproduces it correctly. The $M_{\gamma\gamma}$ 
distribution for data events in which at least one photon is detected
outside the signal-selection range $36^\circ<\theta_\gamma<144^\circ$
is shown in Fig.~\ref{fig:ggg}. This distribution
is dominated by the process $e^+e^-\to3\gamma$. The expected number of
signal events does not exceed 1\% of the number of events in this
distribution, and the expected hadronic background is 0.2\%. Therefore,
the signal contribution is fixed at zero in the fit
($N_{\pi^0\gamma}\equiv0$). The fit result is shown by the histogram.
Good agreement between the simulated and experimental distributions is
observed ($\chi^2/\mathrm{ndf}=61.1/63$, where ndf is the number of
degrees of freedom). The integral of the linear component is
$(-0.5\pm1.1)\%$ of the total number of background events, i.e., it is
consistent with zero. Thus, the simulation correctly reproduces the
shape of the $M_{\gamma\gamma}$ distribution for the dominant
$e^+e^-\to3\gamma$ background.

The $M_{\gamma\gamma}$ spectrum for data events selected with the
standard criteria at $\sqrt{s}=1.282$~GeV is shown in Fig.~\ref{fig:mgg}.
The fit to this distribution yields $N_{\pi^0\gamma}=438\pm80$, with an
acceptable fit quality ($P(\chi^2)=5.7\%$), and the number of background
events $N_{\mathrm{bkg}}=8560\pm124$, where the contribution of the
linear function is $(2.2\pm1.5)\%$ of the total background.

A possible source of systematic uncertainty in $N_{\pi^0\gamma}$ is an
inaccurate description of the background shape. To estimate it, an
alternative fit is performed with the parameters of the linear function
fixed at zero and a free scale factor $\alpha_{\rm{bkg}}$ that
multiplies the expected background. The systematic uncertainty is determined as
the difference in $N_{\pi^0\gamma}$ between the standard and alternative
fits. At $\sqrt{s}=1.282$~GeV, it is 2.7\% with
$\alpha_{\rm{bkg}}=1.02\pm0.02$. The second source is an inaccurate
description of the signal shape. The parameters $\Delta M$ and
$\Delta\sigma^2$, which describe the difference in the signal shape
between data and simulation, were determined at the energy points near
the maximum of the $\phi(1020)$ resonance. Since the energy dependence
of these parameters cannot be determined from our data, the values found
at the $\phi$ meson are used at all energies. The resulting
systematic uncertainty is conservatively estimated as the difference in
$N_{\pi^0\gamma}$ between the fit using $\Delta M$ and $\Delta\sigma^2$
values determined at the $\phi$ meson and the fit with these parameters
set to zero. At
$\sqrt{s}=1.282$~GeV, it is 6.2\%. The total systematic uncertainty in
$N_{\pi^0\gamma}$ is calculated as the quadratic sum of these
contributions.

The $N_{\pi^0\gamma}$ values are determined in all energy intervals.
The results with statistical and systematic uncertainties are listed in
Table~\ref{table}.
\section{Detection efficiency\label{effcor}}

The detection efficiency is determined using simulation of the process
$e^+e^-\to\pi^0\gamma$ as $\varepsilon_0=N^{\rm MC}/N^{\rm MC}_0$, where
$N^{\rm MC}$ is the number of events that pass the selection criteria,
and $N^{\rm MC}_0$ is the total number of generated events. To calculate
the efficiency, events with $E_{\rm ISR}<50$~MeV are used for the
energy points at or below 1.125~GeV, and events with $E_{\rm ISR}<100$~MeV
are used for the energy points above.

The efficiency obtained in this way must be corrected to account for
the imperfect simulation of the detector response. Some of the
systematic uncertainties in the efficiency cancel upon normalization to
the luminosity. These are the uncertainties associated with
the hardware event selection, extra spurious charged tracks
and photons in the event, and the conditions on the parameters 
$E_{\mathrm{tot}}$ and $P_{\mathrm{tot}}$.
The corrected efficiency is determined as
\begin{equation}
\varepsilon=\varepsilon_0\prod_i(1-\delta_i),
\label{eq:eff_corr}
\end{equation}
where $\delta_i$ is the correction for the $i$-th condition.

The correction for the data/simulation difference in photon conversion
before the tracking system does not fully cancel during the
normalization. It is calculated as the difference between the
photon-conversion corrections for 
$e^+e^-\to \pi^0\gamma$~\cite{achasov2025measurement}
and $e^+e^-\to \gamma\gamma$~\cite{achasov2024measurement} and is
equal to $\delta_{\rm conv}=(0.27\pm0.03)\%$.
The correction for the $\chi^2_{3\gamma}$
condition ($\delta_{\chi^2}=(-0.9\pm0.7)\%$) was obtained in
Ref.~\cite{achasov2025measurement}. 

We do not introduce corrections for the conditions on $\theta_\gamma$
and $E_{\gamma,\mathrm{min}}$. Instead, we estimate the associated
systematic uncertainty in the following way. The QED process
$e^+e^-\to 3\gamma$ is very sensitive to the conditions on $\theta_\gamma$
and $E_{\gamma,\mathrm{min}}$. 
In the fit to the $M_{\gamma\gamma}$ spectrum in Sec.~\ref{mrec}
we determine, in particular, the scale factor for the calculated 
$3\gamma$ background. The energy dependence of the scale factors 
obtained for all energy intervals
is consistent with a constant. Its average value is $1.028\pm0.007$.
This value should be corrected for $\delta_{\rm conv}$ and
$\delta_{\chi^2}$.
The final number is $1.022\pm0.017$, where the uncertainty includes
the statistical error, uncertainties in the efficiency corrections
(0.7\%), integrated luminosity (1\%), and the theoretical uncertainty
of the $e^+e^-\to 3\gamma$ cross section, which is estimated to be 
about 1\%. The average corrected scale factor is consistent with
unity. We take the quadratic sum of its deviation from unity and
its uncertainty (2.8\%) as a conservative estimate of the uncertainty
associated with the conditions on $\theta_\gamma$ and 
$E_{\gamma,\mathrm{min}}$.

The total correction is
\begin{equation*}
\prod_i(1-\delta_i)=1.006\pm0.029,
\end{equation*}
where the quoted error is the total systematic uncertainty of the
detection efficiency, which includes the uncertainty of the correction
and the uncertainty
associated with the conditions on $\theta_\gamma$ and
$E_{\gamma,\mathrm{min}}$. The corrected efficiency values are listed in
Table~\ref{table}.

The twofold decrease in the efficiency in the energy region under
study (see Table~\ref{table}) has two causes. The first is the change
in the condition on $E_{\gamma,\mathrm{min}}$ at $\sqrt{s}=1.625$~GeV,
which leads to a 16\% decrease. The second is the merging
of the electromagnetic showers from the photons of the $\pi^0$ decay.
The minimum opening angle between these photons changes from 
$28^\circ$ at $\sqrt{s}=1.1$~GeV to $15^\circ$ at $\sqrt{s}=2.0$~GeV,
while the angular size of a calorimeter crystal is $9^\circ$.

\begin{figure}[t!]
\includegraphics[width=0.47\textwidth]{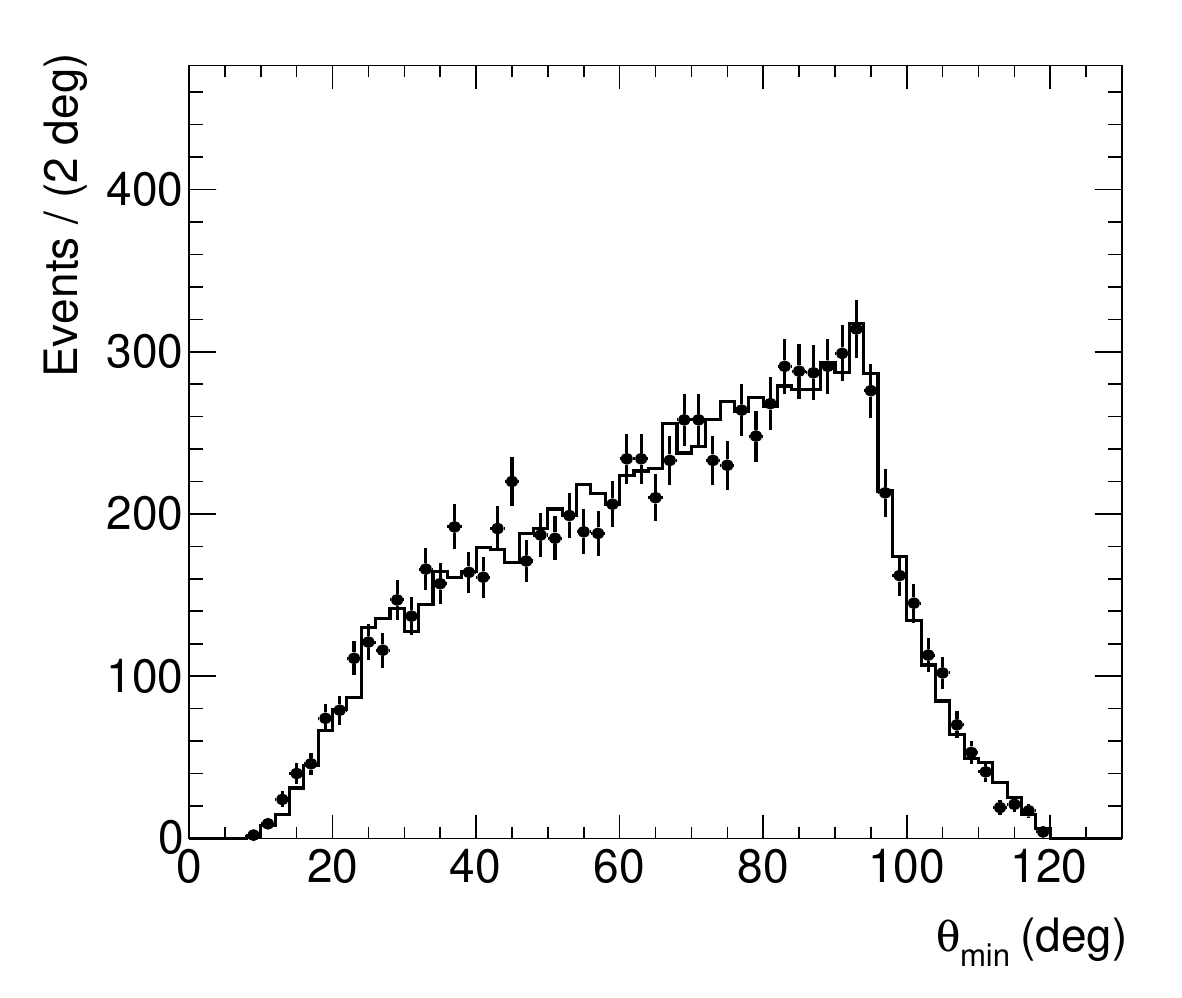}
\includegraphics[width=0.47\textwidth]{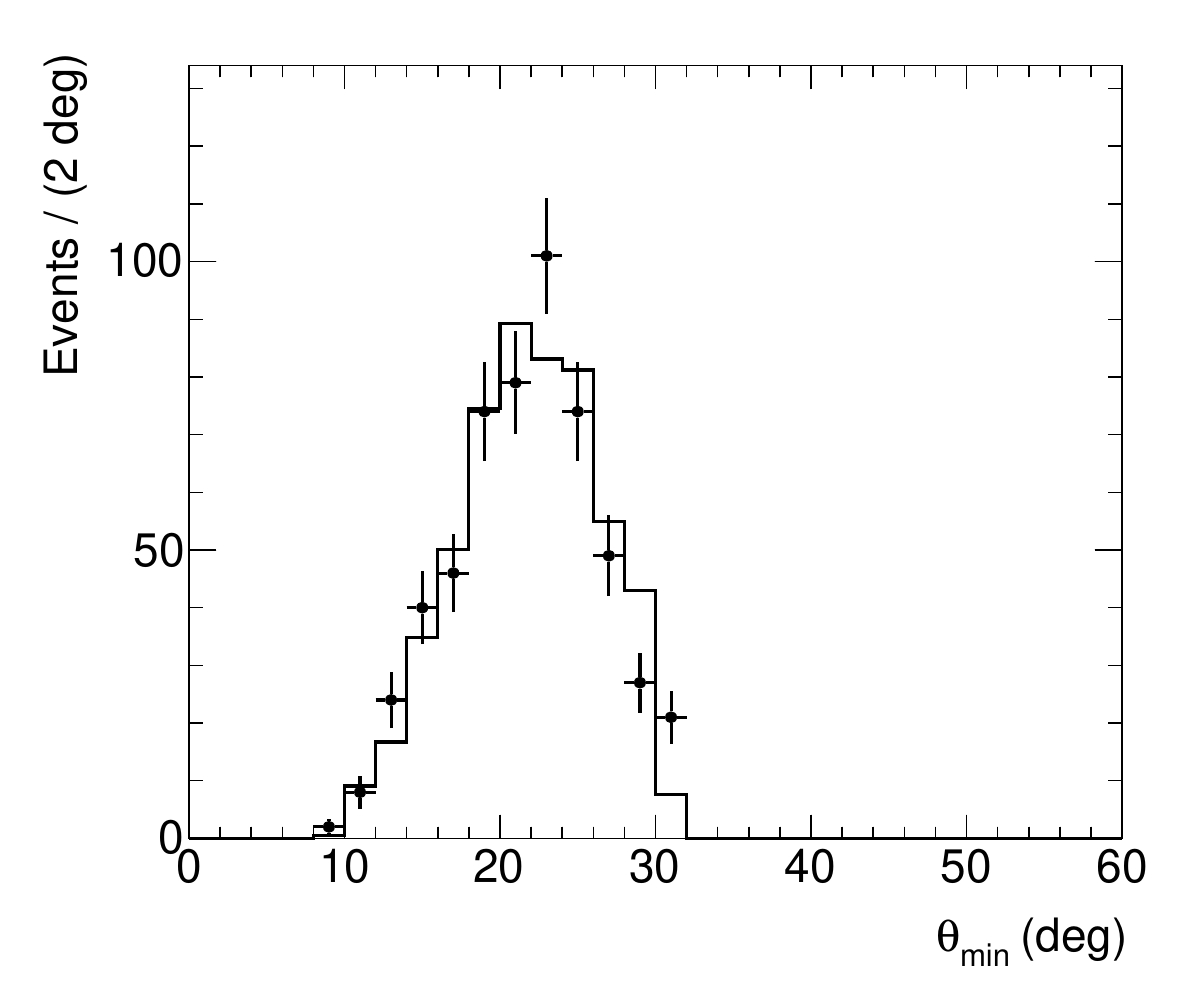}
\caption{The distribution of the minimum opening angle between photons at
$\sqrt{s}=2.0$~GeV for data (points with error bars) and simulated
events (histogram). Left panel: events that pass the standard selection criteria
without the condition $70<M_{\gamma\gamma}<200$~MeV. Right panel: events
that pass the standard selection criteria.
\label{fig:minAng2000}}
\end{figure}

The distributions of the minimum opening angle between the reconstructed
photons at $\sqrt{s}=2.0$~GeV, where the $\pi^0\gamma$ contribution is
negligible compared to the background from the process
$e^+e^-\to3\gamma$, are shown in Fig.~\ref{fig:minAng2000}. The left
panel shows the distribution for events that pass the standard selection
criteria except for the condition $70<M_{\gamma\gamma}<200$~MeV. The
position and shape of the left edge of this distribution are determined
by the ability of the event reconstruction algorithm to separate close
photons. The right panel shows the same distribution with the condition
$70<M_{\gamma\gamma}<200$~MeV applied. The position of the right edge of
this distribution is determined by the condition
$E_{\gamma,\mathrm{min}}>150$~MeV. It is evident that both experimental
distributions agree reasonably well with the simulation. This indicates
that the shape of photon showers and the reconstruction of close photons
are simulated correctly.
\section{Determination of the Born cross section\label{fullcrs}}

The visible cross section determined directly from the experimental
data is
\begin{equation}\label{f30}
\sigma_{\mathrm{exp},i}=\frac{N_{\pi^0\gamma,i}}{\varepsilon_i\,\mathrm{IL}_i},
\end{equation}
where $i$ is the index of the energy interval. The visible cross
section is related to the Born cross section $\sigma_{\mathrm{Born}}$ as
follows~\cite{kuraev1985radiative}:
\begin{equation}\label{f31}
\sigma_{\rm exp}(s)=\int_0^{x_{\max}}F(x,s)\,
\sigma_{\rm Born}\!\big(s(1-x)\big)\,dx=\sigma_{\rm Born}(s)\,(1+\delta(s)),
\end{equation}
where $F(x,s)$ is a radiator function that describes the probability of
emitting photons with total energy $E_{\mathrm{ISR}}$ from the initial
state, $x=2E_{\mathrm{ISR}}/\sqrt{s}$, and $\delta(s)$ is the radiative
correction. The upper limit of integration is
$x_{\max}=2E_{\mathrm{ISR,max}}/\sqrt{s}$, where
$E_{\mathrm{ISR,max}}=50$~MeV for $\sqrt{s}\le 1.125$~GeV and 100~MeV for
$\sqrt{s}>1.125$~GeV.

The energy dependence of the measured visible cross section is fitted
with Eq.~(\ref{f31}), in which $\sigma_{\rm Born}(s)$ is described by a
theoretical model. The fit result is used to calculate the radiative
correction $\delta(s_i)$ and the experimental values of the Born cross
section $\sigma_{{\rm Born},i}=\sigma_{{\rm exp},i}/(1+\delta(s_i))$.

In this work, the Born cross section is described within the vector
meson dominance (VMD) model~\cite{o1997rho}:
\begin{equation}\label{f56}
\sigma_{\rm{Born}}(s)=\frac{q^3(s)}{s^{3/2}}
\left|\sum_V A_V(s)\right|^2=\frac{q^3(s)}{s^{3/2}}
\left|A_\rho(s)+A_\omega(s)+A_\phi(s)+A_{V_1}(s)+A_{V_2}(s)\right|^2,
\end{equation}
\begin{equation}
A_V(s)=\frac{m_V\Gamma_V e^{i\varphi_V}}
{m_V^2-s-i\sqrt{s}\,\Gamma_V(s)}
\sqrt{\frac{m_V^3}{q^3(m_V^2)}\,\sigma_V}, \quad
\sigma_V=\frac{12\pi}{m_V^2}\mathcal{B}(V\to e^+e^-)\mathcal{B}(V\to\pi^0\gamma),
\label{f56a}
\end{equation}
where $q(s)=\tfrac{\sqrt{s}}{2}(1-m_{\pi^0}^2/s)$ is the $\pi^0$
momentum, $m_V$ is the mass of the vector resonance, $\Gamma_V(s)$ is
its energy-dependent width ($\Gamma_V\equiv\Gamma_V(m_V^2)$), $\varphi_V$
is the interference phase, and $\sigma_V$ is the cross section at the
resonance maximum. The sum in Eq.~(\ref{f56}) runs over the
$\rho(770)$, $\omega(782)$, and $\phi(1020)$ resonances and the excited
states of the $\rho$ and $\omega$ families.

Since the resonance pairs $\omega(1420)$--$\rho(1450)$ and
$\omega(1650)$--$\rho(1700)$ have close masses and widths, their
contributions cannot be resolved in the fit to the cross section.
Therefore, we introduce two effective resonances, $V_1$ and $V_2$, which
describe the total contributions of these pairs, respectively,
including the interference within each pair. To calculate the line
shape of an effective resonance, we use the quark model relation that
the products $\Gamma(V\to e^+e^-)\Gamma(V\to\pi^0\gamma)$ are equal for
the $\rho$ and $\omega$ partners within each pair, with their amplitudes
adding coherently with a relative phase of $0^\circ$. The
parameters of the $\omega(1420)$ resonance are currently poorly known
(see Particle Data Group (PDG) tables~\cite{pdg2026}). Therefore, 
the mass and width of the
$\rho(1450)$ are used for $V_1$: $m_{V_1}=1465\pm25$~MeV and
$\Gamma_{V_1}=400\pm60$~MeV~\cite{pdg2026}. The parameters of $V_2$ are
determined by fitting the coherent sum of the $\omega(1650)$
($m=1670$~MeV, $\Gamma=150$~MeV~\cite{pdg2026}) and $\rho(1700)$
($m=1790$~MeV, $\Gamma=300$~MeV~\cite{pdg2026}) amplitudes with a
single resonance of the form~(\ref{f56a}). As a result,
$m_{V_2}=1680\pm35$~MeV and $\Gamma_{V_2}=210\pm80$~MeV are obtained,
where the uncertainties reflect the spread of the $\omega(1650)$ and
$\rho(1700)$ parameters.

The masses and widths of the $\rho(770)$, $\omega(782)$, and
$\phi(1020)$ are fixed at their PDG values~\cite{pdg2026}, as are
the branching fractions of the $\rho(770)$ and $\omega(782)$.
In the calculation of the energy dependence of the $\rho(770)$,
$\omega(782)$, and $\phi(1020)$ widths, all decay modes with branching
fractions above 1\% are taken into account. For the effective
resonances $V_1$ and $V_2$, the total widths are assumed to be energy
independent. The phase of the $\omega(782)$ amplitude is set to zero,
and the phases of the $\rho(770)$ and $\phi(1020)$ are fixed at
$-12.7^\circ$~\cite{achasov2016study} and
$163^\circ$~\cite{achasov2003study}, respectively. For the effective
resonances $V_1$ and $V_2$, the expected phases are close to $180^\circ$
and $0^\circ$~\cite{clegg1994higher}. The value $\varphi_{V_1}\approx180^\circ$
for the $\omega(1420)$ and $\rho(1450)$ resonances is confirmed by the
cross section measurements in the $e^+e^-\to\rho\pi$~\cite{aul2015study}
and $\omega\pi^0$~\cite{achasov2013study} channels. Therefore,
$\varphi_{V_1}$ is fixed at $180^\circ$. For the phase $\varphi_{V_2}$,
two values, $0^\circ$ and $180^\circ$, are considered.

\begin{table}
\caption{The c.m. energy ($\sqrt{s}$), integrated luminosity
($\mathrm{IL}$), detection efficiency ($\varepsilon$), number of selected
$e^+e^-\to\pi^0\gamma$ events ($N_{\pi^0\gamma}$), radiative correction
($1+\delta$), and measured Born cross section ($\sigma_{\rm{Born}}$).
For the number of events and the cross section, the first uncertainty
is statistical and the second is systematic. For the radiative
correction, the quoted uncertainty is systematic.
The energy range of each interval is given in parentheses.\label{table}}
\footnotesize
\begin{ruledtabular}
\begin{tabular}{cccccc}
$\sqrt{s}$ (GeV) & $\mathrm{IL}$ (pb$^{-1}$) & $\varepsilon$ (\%) & $N_{\pi^0\gamma}$ & $1+\delta$ & $\sigma_{\rm Born}$ (pb) \\
\hline
1.100 (1.080--1.125) & 17.89 & 27.6 & 128 $\pm$ 50 $\pm$ 9 & 0.892 $\pm$ 0.001 & 29 $\pm$ 11 $\pm$ 2 \\
1.160 (1.140--1.180) & 22.44 & 27.6 & 211 $\pm$ 54 $\pm$ 11 & 0.955 $\pm$ 0.003 & 36 $\pm$ 9 $\pm$ 2 \\
1.220 (1.200--1.240) & 20.87 & 28.5 & 133 $\pm$ 50 $\pm$ 9 & 0.959 $\pm$ 0.004 & 23 $\pm$ 9 $\pm$ 2 \\
1.267 (1.250--1.276) & 20.88 & 28.3 & 178 $\pm$ 48 $\pm$ 7 & 0.952 $\pm$ 0.004 & 32 $\pm$ 9 $\pm$ 2 \\
1.282 (1.282--1.282) & 62.73 & 27.8 & 438 $\pm$ 80 $\pm$ 30 & 0.950 $\pm$ 0.005 & 26 $\pm$ 5 $\pm$ 2 \\
1.317 (1.300--1.340) & 25.06 & 28.4 & 86 $\pm$ 50 $\pm$ 6 & 0.943 $\pm$ 0.006 & 13 $\pm$ 7 $\pm$ 1 \\
1.374 (1.350--1.400) & 26.65 & 28.0 & 129 $\pm$ 66 $\pm$ 5 & 0.934 $\pm$ 0.010 & 19 $\pm$ 9 $\pm$ 1 \\
1.445 (1.420--1.460) & 26.64 & 27.4 & 97 $\pm$ 47 $\pm$ 2 & 0.924 $\pm$ 0.016 & 14 $\pm$ 7 $\pm$ 1 \\
1.500 (1.475--1.525) & 21.45 & 26.9 & 70 $\pm$ 41 $\pm$ 0 & 0.918 $\pm$ 0.016 & 13 $\pm$ 8 $\pm$ 0 \\
1.552 (1.540--1.560) & 17.58 & 26.4 & 31 $\pm$ 39 $\pm$ 8 & 0.911 $\pm$ 0.014 & 7 $\pm$ 9 $\pm$ 2 \\
1.578 (1.575--1.580) & 28.49 & 26.0 & 52 $\pm$ 45 $\pm$ 2 & 0.907 $\pm$ 0.018 & 8 $\pm$ 7 $\pm$ 0 \\
1.601 (1.594--1.620) & 22.96 & 25.4 & 39 $\pm$ 39 $\pm$ 1 & 0.906 $\pm$ 0.025 & 7 $\pm$ 7 $\pm$ 0 \\
1.649 (1.625--1.675) & 18.52 & 21.2 & 49 $\pm$ 30 $\pm$ 2 & 0.914 $\pm$ 0.049 & 14 $\pm$ 8 $\pm$ 1 \\
1.680 (1.680--1.680) & 21.18 & 20.4 & 49 $\pm$ 32 $\pm$ 2 & 0.935 $\pm$ 0.063 & 12 $\pm$ 8 $\pm$ 1 \\
1.719 (1.700--1.740) & 22.71 & 19.6 & 42 $\pm$ 31 $\pm$ 3 & 0.982 $\pm$ 0.097 & 10 $\pm$ 7 $\pm$ 1 \\
1.772 (1.750--1.780) & 21.43 & 18.9 & 9 $\pm$ 29 $\pm$ 5 & 1.044 $\pm$ 0.203 & 2 $\pm$ 7 $\pm$ 1 \\
1.823 (1.800--1.840) & 27.57 & 18.2 & 18 $\pm$ 33 $\pm$ 5 & 1.057 $\pm$ 0.198 & 3 $\pm$ 7 $\pm$ 1 \\
1.869 (1.850--1.875) & 65.47 & 17.0 & 6 $\pm$ 49 $\pm$ 7 & 1.041 $\pm$ 0.173 & 1 $\pm$ 4 $\pm$ 1 \\
1.877 (1.877--1.878) & 26.31 & 16.3 & 66 $\pm$ 32 $\pm$ 9 & 1.038 $\pm$ 0.171 & 15 $\pm$ 7 $\pm$ 3 \\
1.880 (1.879--1.882) & 38.19 & 16.2 & 32 $\pm$ 38 $\pm$ 4 & 1.036 $\pm$ 0.171 & 5 $\pm$ 6 $\pm$ 1 \\
1.895 (1.884--1.910) & 106.8 & 16.6 & 55 $\pm$ 77 $\pm$ 27 & 1.029 $\pm$ 0.171 & 3 $\pm$ 4 $\pm$ 2 \\
1.934 (1.920--1.950) & 34.82 & 16.1 & $-62$ $\pm$ 36 $\pm$ 2 & 1.011 $\pm$ 0.181 & $-11$ $\pm$ 6 $\pm$ 2 \\
1.971 (1.960--1.980) & 22.35 & 15.0 & $-29$ $\pm$ 27 $\pm$ 3 & 0.995 $\pm$ 0.208 & $-9$ $\pm$ 8 $\pm$ 2 \\
2.000 (2.000--2.000) & 25.75 & 13.9 & $-15$ $\pm$ 29 $\pm$ 18 & 0.983 $\pm$ 0.239 & $-4$ $\pm$ 8 $\pm$ 5 \\
\end{tabular}
\end{ruledtabular}
\end{table}

\begin{figure}
\includegraphics[width=\textwidth]{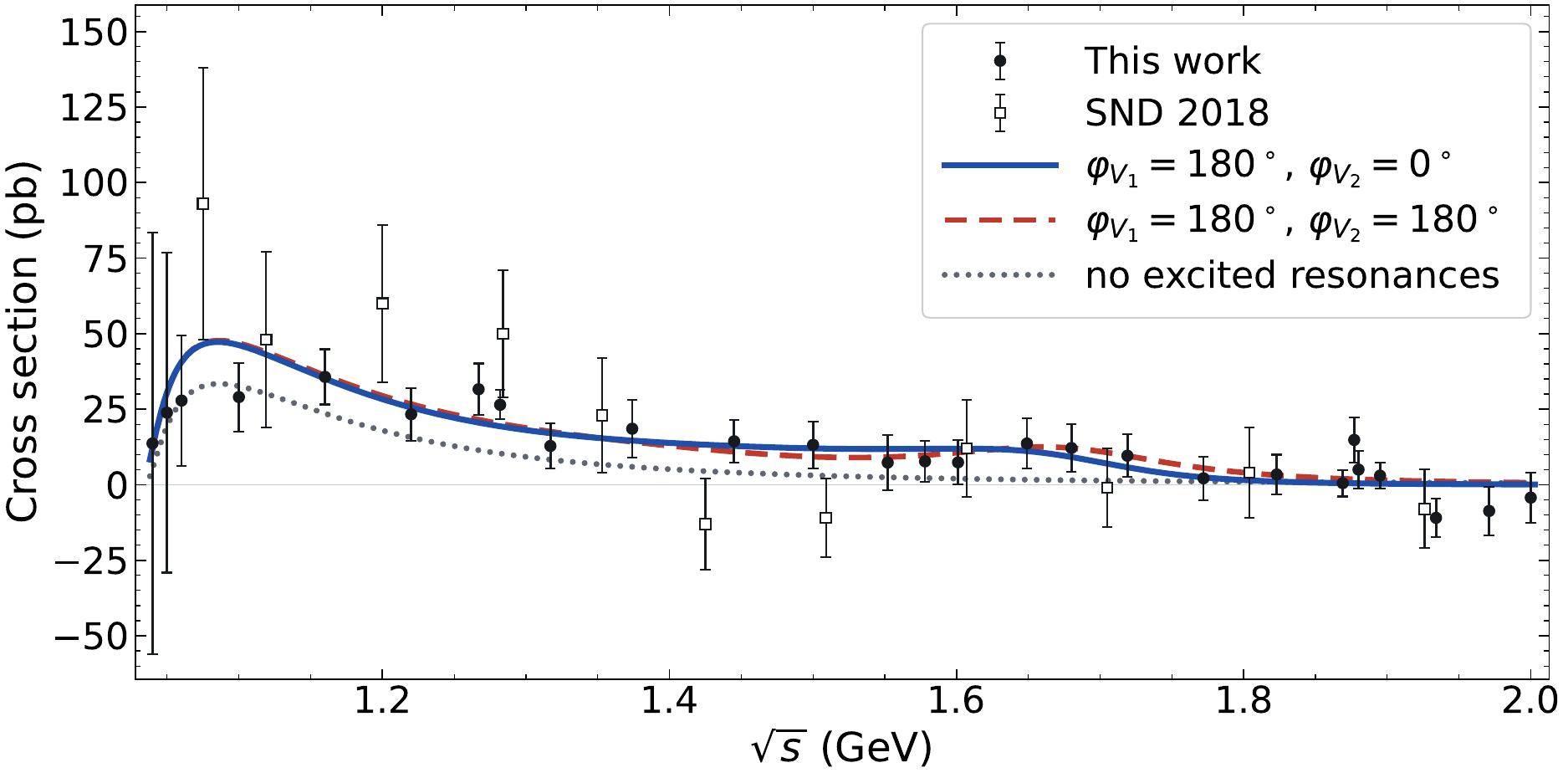}
\caption{Comparison of the $e^+e^-\to\pi^0\gamma$ Born cross section
measured in this work (points with error bars) with the results of the
previous most accurate measurement~\cite{achasov2018measurement}
(squares). The solid and dashed curves are the results of the fit to the
data in models 3 and 2, respectively. The dotted curve is the null
hypothesis without the contribution of the excited resonances.
\label{fig:crs}}
\end{figure}

The cross section measured in this work is fitted together with the
data from Ref.~\cite{achasov2025measurement} obtained in
the $\phi(1020)$ resonance region (1.000--1.060~GeV). This makes it
possible to study how the inclusion of the higher-lying resonances
affects the value of $\mathcal{B}(\phi\to\pi^0\gamma)$. The only free
fit parameters are $\sigma_{V_1}$, $\sigma_{V_2}$, and
$\mathcal{B}(\phi\to\pi^0\gamma)$. All other parameters are fixed as
described above, and two values of $\varphi_{V_2}$ are
considered.

\begin{table}
\caption{The values of $\chi^2/\rm{ndf}$, the peak cross sections of
the effective resonances, and the branching fraction
$\mathcal{B}(\phi\to\pi^0\gamma)$ obtained from the simultaneous fit to
the data from this work and Ref.~\cite{achasov2025measurement} for three
models. The $\chi^2$ value is calculated using only the data from this
work. For the data in the $\phi(1020)$ resonance region,
$\chi^2/\rm{ndf}=17/16$ for all three models. For model 3, the first
uncertainty in the cross sections is statistical, and the second is the
model uncertainty due to the uncertainties in the masses and widths of
the effective resonances.\label{table2}}
\begin{ruledtabular}
\begin{tabular}{cccccc}
Model & ($\varphi_{V_1},\varphi_{V_2}$) & $\chi^2/\rm{ndf}$ & $\sigma_{V_1}$ (pb) & $\sigma_{V_2}$ (pb) & $\mathcal{B}(\phi\to\pi^0\gamma)$ ($10^{-3}$) \\
\hline
1 & $-,\ -$                 & $41/24$ & $0$           & $0$          & $1.484\pm0.017$ \\
2 & $180^\circ,\ 180^\circ$ & $16/22$ & $4^{+5}_{-3}$ & $11\pm6$     & $1.434\pm0.022$ \\
3 & $180^\circ,\ 0^\circ$   & $18/22$ & $32\pm5\pm8$  & $19\pm7\pm7$ & $1.416\pm0.018$ \\
\end{tabular}
\end{ruledtabular}
\end{table}

The fit results for the two values of the phase $\varphi_{V_2}$ and for
the model without excited resonances (null hypothesis) are listed in
Table~\ref{table2}. It is evident that the fit quality is good for the
models with $\varphi_{V_2}=0^\circ$ and $180^\circ$. The model with the
expected phase value $\varphi_{V_2}=0^\circ$~\cite{clegg1994higher} is
chosen to calculate the radiative corrections. The $\chi^2$ difference
between the null hypothesis and this model corresponds to a significance
of the excited-resonance contribution of $4.4\sigma$.

The calculated radiative corrections and the Born cross section are
listed in Table~\ref{table}. The sources of systematic uncertainty in
the cross section are the inaccuracies in the description of the
signal and background shapes, in the simulation of the detection
efficiency (2.9\%), in the integrated luminosity (1.0\%), and in the
radiative correction. The systematic uncertainty of the radiative
correction is estimated by varying the free model parameters obtained
from the fit within their errors, as well as by varying the masses and
widths of the effective resonances within the limits given above. The
model uncertainty, estimated as the difference between the radiative
corrections in models 2 and 3 (Table~\ref{table2}), is added to this
uncertainty in quadrature.

The measured cross section and the fit results are shown in
Fig.~\ref{fig:crs}, together with the data from the previous most accurate
measurement~\cite{achasov2018measurement}, also performed with the SND
detector. The cross section obtained in this work
is consistent with the previous measurement, while its statistical accuracy is
improved by a factor of 4.

In Ref.~\cite{achasov2025measurement}, devoted to the measurement of
the $e^+e^-\to\pi^0\gamma$ cross section near the $\phi(1020)$
resonance, the contribution of the higher-lying resonances was
described by a single effective resonance $V_1$ with a mass and width
close to those of the $\rho(1450)$. Its peak cross section was not
constrained by data and was a free parameter in the fit to
the data in the $\phi(1020)$ resonance region. The data obtained in
this work constrain the contribution of the higher-lying resonances in
the $\phi(1020)$ region and make it possible to refine the value of
$\mathcal{B}(\phi\to\pi^0\gamma)$. Our result obtained in model 3,
\begin{equation}
\mathcal{B}(\phi\to\pi^0\gamma)=(1.416\pm0.018\pm0.071)\times10^{-3},
\end{equation}
supersedes the result of Ref.~\cite{achasov2025measurement},
$\mathcal{B}(\phi\to\pi^0\gamma)=(1.452\pm0.024\pm0.067)\times10^{-3}$.
The systematic uncertainty of the new value includes the model
uncertainty of $0.023\times10^{-3}$, which is obtained by
combining in quadrature the difference between the 
$\mathcal{B}(\phi\to\pi^0\gamma)$ values in models 2 and 3
($0.018\times10^{-3}$) and the change in
$\mathcal{B}(\phi\to\pi^0\gamma)$ when the masses and widths of the
effective resonances are varied ($0.014\times10^{-3}$).

The peak cross sections from Table~\ref{table2} can be compared with the
predictions of the quark model~\cite{close2002radiative}. The peak cross
section of a resonance is related to its total production cross section
as $\sigma_{V\to\pi^0\gamma}=\sigma^V_{\rm tot}\,\Gamma_{\pi^0\gamma}/\Gamma_V$.
Estimating $\sigma^V_{\rm tot}$ from experimental data at
$\sqrt{s}=m_V$ as the sum of the cross sections for the main final
states, we obtain approximately 60~nb for the $\rho(1450)$
($\pi^+\pi^-\pi^0\pi^0$~\cite{babar2017fourpi},
$\pi^+\pi^-\pi^+\pi^-$~\cite{babar2012fourpi}, and
$\eta\pi^+\pi^-$~\cite{achasov2018etapipi}) and 8~nb for the
$\omega(1650)$ ($\pi^+\pi^-\pi^0$~\cite{achasov2024threepi},
$\omega\pi\pi$~\cite{babar2007fivepi}, and
$\omega\eta$~\cite{achasov2016omegaeta}). With the total widths
$\Gamma_{\rho(1450)}=400$~MeV and $\Gamma_{\omega(1650)}=150$~MeV~\cite{pdg2026}
and the $\pi^0\gamma$ decay widths of 61~keV for the $\rho(1450)$ and
40~keV for the $\omega(1650)$~\cite{close2002radiative}, this gives
$\sigma_{\rho(1450)\to\pi^0\gamma}\approx10$~pb and
$\sigma_{\omega(1650)\to\pi^0\gamma}\approx2$~pb. Using these values and
the procedure for determining the line shape of the effective resonances
described above, we obtain
\begin{equation}
\sigma_{V_1}\approx40\mbox{ pb}, \qquad
\sigma_{V_2}\approx4\mbox{ pb}.
\end{equation}
The accuracy of the theoretical predictions for the $\pi^0\gamma$ decay
widths of the excited resonances is estimated in
Ref.~\cite{close2002radiative} to be 50\%. The estimate for
$\sigma_{V_1}$ agrees well with the fit result in model 3,
$\sigma_{V_1}=32\pm5\pm8$~pb. The measured value
$\sigma_{V_2}=19\pm7\pm7$~pb, although it exceeds the prediction by
about a factor of 5, is consistent with it within the uncertainties
(the difference is $1.5\sigma$).
\section{Summary}

In the experiment with the SND detector at the VEPP-2000 $e^+e^-$
collider, the process $e^+e^-\to\pi^0\gamma$ has been studied in the
c.m. energy range 1.08--2.0~GeV. The analysis is based on data with an
integrated luminosity of 745~pb$^{-1}$, accumulated in 2017--2024. The
most accurate measurement to date of the $e^+e^-\to\pi^0\gamma$ cross
section as a function of the c.m. energy has been performed. Compared to the previous most accurate measurement~\cite{achasov2018measurement},
the statistical accuracy is improved by a factor of 4, with the 
measurement still statistically limited. The energy
dependence of the cross section is well described by the vector meson
dominance model, in which the contribution of the excited vector states
is described by two effective resonances. The model without excited
resonances is excluded at the level of $4.4\sigma$.


\begin{thebibliography}{99}
\bibitem{landsberg1985electromagnetic} L.~G.~Landsberg,
Phys.\ Rep.\ {\bf 128}, 301 (1985).
\bibitem{o1981radiative} P.~J.~O'Donnell,
Rev.\ Mod.\ Phys.\ {\bf 53}, 673 (1981).
\bibitem{o1997rho} H.~B.~O'Connell, B.~C.~Pearce, A.~W.~Thomas and A.~G.~Williams,
Prog.\ Part.\ Nucl.\ Phys.\ {\bf 39}, 201 (1997).
\bibitem{pdg2026} F.~Takahashi {\it et al.} (Particle Data Group),
Int.\ J.\ Mod.\ Phys.\ A {\bf 41}, 2630011 (2026).
\bibitem{isgur1985gluonic} N.~Isgur, R.~Kokoski and J.~Paton,
Phys.\ Rev.\ Lett.\ {\bf 54}, 869 (1985).
\bibitem{close1995page} F.~E.~Close and P.~R.~Page,
Nucl.\ Phys.\ B {\bf 443}, 233 (1995).
\bibitem{close2002radiative} F.~E.~Close, A.~Donnachie and Yu.~S.~Kalashnikova,
Phys.\ Rev.\ D {\bf 65}, 092003 (2002).
\bibitem{jaffe1977} R.~L.~Jaffe,
Phys.\ Rev.\ D {\bf 15}, 281 (1977).
\bibitem{aoyama2020} T.~Aoyama {\it et al.},
Phys.\ Rep.\ {\bf 887}, 1 (2020).
\bibitem{aliberti2025} R.~Aliberti {\it et al.},
Phys.\ Rep.\ {\bf 1143}, 1 (2025).
\bibitem{Hoferichter2018} M.~Hoferichter, B.-L.~Hoid, B.~Kubis, S.~Leupold and S.~P.~Schneider,
J.\ High Energy Phys.\ {\bf 10}, 141 (2018).
\bibitem{achasov2016study} M.~N.~Achasov {\it et al.} (SND Collaboration),
Phys.\ Rev.\ D {\bf 93}, 092001 (2016).
\bibitem{achasov2018measurement} M.~N.~Achasov {\it et al.} (SND Collaboration),
Phys.\ Rev.\ D {\bf 98}, 112001 (2018).
\bibitem{achasov2025measurement} M.~N.~Achasov {\it et al.} (SND Collaboration),
Phys.\ Atom.\ Nucl.\ {\bf 88}, 453 (2025).
\bibitem{akhmetshin2005study} R.~R.~Akhmetshin {\it et al.} (CMD-2 Collaboration),
Phys.\ Lett.\ B {\bf 605}, 26 (2005).
\bibitem{shatunov2016status} P.~Yu.~Shatunov {\it et al.},
Phys.\ Part.\ Nucl.\ Lett.\  {\bf 13}, 995 (2016).
\bibitem{achasov2000spherical} M.~N.~Achasov {\it et al.},
Nucl.\ Instrum.\ Methods Phys.\ Res., Sect.\ A {\bf 449}, 125 (2000).
\bibitem{achasov2009first} M.~N.~Achasov {\it et al.},
Nucl.\ Instrum.\ Methods Phys.\ Res., Sect.\ A {\bf 598}, 31 (2009).
\bibitem{aulchenko2009snd} V.~M.~Aulchenko {\it et al.},
Nucl.\ Instrum.\ Methods Phys.\ Res., Sect.\ A {\bf 598}, 102 (2009).
\bibitem{barnyakov2009high} A.~Yu.~Barnyakov {\it et al.},
Nucl.\ Instrum.\ Methods Phys.\ Res., Sect.\ A {\bf 598}, 163 (2009).
\bibitem{aulchenko2009daq} V.~M.~Aulchenko {\it et al.},
Nucl.\ Instrum.\ Methods Phys.\ Res., Sect.\ A {\bf 598}, 340 (2009).
\bibitem{timoshenko2020status} M.~V.~Timoshenko {\it et al.},
Phys.\ Part.\ Nucl.\ Lett.\ {\bf 17}, 419 (2020).
\bibitem{kuraev1985radiative} E.~A.~Kuraev and V.~S.~Fadin,
Sov.\ J.\ Nucl.\ Phys.\ {\bf 41}, 466 (1985)
[Yad.\ Fiz.\ {\bf 41}, 733 (1985)].
\bibitem{bonneau1971hard} G.~Bonneau and F.~Martin,
Nucl.\ Phys.\ B {\bf 27}, 381 (1971).
\bibitem{agostinelli2003} S.~Agostinelli {\it et al.} (GEANT4 Collaboration),
Nucl.\ Instrum.\ Methods Phys.\ Res., Sect.\ A {\bf 506}, 250 (2003).
\bibitem{allison2016} J.~Allison {\it et al.},
Nucl.\ Instrum.\ Methods Phys.\ Res., Sect.\ A {\bf 835}, 186 (2016).
\bibitem{achasov2024measurement} M.~N.~Achasov {\it et al.} (SND Collaboration),
Phys.\ Rev.\ D {\bf 110}, 072001 (2024).
\bibitem{balossini2006matching} G.~Balossini, C.~M.~Carloni Calame, G.~Montagna, O.~Nicrosini and F.~Piccinini,
Nucl.\ Phys.\ B {\bf 758}, 227 (2006).
\bibitem{balossini2008photon} G.~Balossini, C.~Bignamini, C.~M.~Carloni Calame, G.~Montagna, O.~Nicrosini and F.~Piccinini,
Phys.\ Lett.\ B {\bf 663}, 209 (2008).
\bibitem{generator2026} A.~A.~Korol, Phys.\ Atom.\ Nucl.\ (to be published),
\url{https://indico.inp.nsk.su/event/140/contributions/3465/attachments/1883/2816/session-korol-2026-03-11.pdf}.
\bibitem{achasov2003study} M.~N.~Achasov {\it et al.} (SND Collaboration),
Phys.\ Rev.\ D {\bf 68}, 052006 (2003).
\bibitem{clegg1994higher} A.~B.~Clegg and A.~Donnachie,
Z.\ Phys.\ C {\bf 62}, 455 (1994).
\bibitem{aul2015study} V.~M.~Aulchenko {\it et al.} (SND Collaboration),
J.\ Exp.\ Theor.\ Phys.\ {\bf 121}, 27 (2015).
\bibitem{achasov2013study} M.~N.~Achasov {\it et al.} (SND Collaboration),
Phys.\ Rev.\ D {\bf 88}, 054013 (2013).
\bibitem{babar2017fourpi} J.~P.~Lees {\it et al.} (BABAR Collaboration),
Phys.\ Rev.\ D {\bf 96}, 092009 (2017).
\bibitem{babar2012fourpi} J.~P.~Lees {\it et al.} (BABAR Collaboration),
Phys.\ Rev.\ D {\bf 85}, 112009 (2012).
\bibitem{achasov2018etapipi} M.~N.~Achasov {\it et al.} (SND Collaboration),
Phys.\ Rev.\ D {\bf 97}, 012008 (2018).
\bibitem{achasov2024threepi} M.~N.~Achasov {\it et al.} (SND Collaboration),
Phys.\ Atom.\ Nucl.\ {\bf 87}, 747 (2024).
\bibitem{babar2007fivepi} B.~Aubert {\it et al.} (BABAR Collaboration),
Phys.\ Rev.\ D {\bf 76}, 092005 (2007).
\bibitem{achasov2016omegaeta} M.~N.~Achasov {\it et al.} (SND Collaboration),
Phys.\ Rev.\ D {\bf 94}, 092002 (2016).
\end{thebibliography}
\end{document}